\documentclass{article}
\usepackage{amssymb}
\usepackage{amsfonts}

\newtheorem{theorem}{Theorem}
\newtheorem{acknowledgement}[theorem]{Acknowledgement}

\input{tcilatex}
\begin{document}

\title{Reply to a comment on "Quantum theory and the limits of objectivity"}
\author{Richard Healey\thanks{%
Philosophy Department, University of Arizona, Tucson, AZ 85721, USA and
Stellenbosch Institute for Advanced Study (STIAS), Wallenberg Research
Centre at Stellenbosch University, Stellenbosch 7600, South Africa.} \\
}
\date{}
\maketitle

\begin{abstract}
In this short note I reply to criticisms of an argument in my paper \cite{1}
that appear in comment \cite{2}. I refer the reader to section 4 of \cite{1}
in which I described the scenario of a \textit{Gedankenexperiment} on which
is based the argument criticized in \cite{2}. The authors of \cite{2} raise
one \textquotedblleft main criticism\textquotedblright\ then go on to claim
that the argument of \cite{1} contains a series of problems. But their
\textquotedblleft main criticism\textquotedblright\ is not an objection to
the argument and the problems are of their own making. In replying to what
they call their main criticism I will take this opportunity to exhibit the
structure of the argument and so make clear why this is not an objection to
that argument.
\end{abstract}

In this short note I reply to criticisms of an argument in my paper \cite{1}
that appear in comment \cite{2}. I refer the reader to section 4 of \cite{1}
in which I described the scenario of a \textit{Gedankenexperiment} on which
is based the argument criticized in \cite{2}. The authors of \cite{2} raise
one \textquotedblleft main criticism\textquotedblright\ then go on to claim
that the argument of \cite{1} contains a series of problems. But their
\textquotedblleft main criticism\textquotedblright\ is not an objection to
the argument and the problems are of their own making. In replying to what
they call their main criticism I will take this opportunity to exhibit the
structure of the argument and so make clear why this is not an objection to
that argument.

I begin by addressing the \textquotedblleft main
criticism\textquotedblright\ offered in \cite{2}, there stated as follows:

\textquotedblleft \ldots the computed correlation functions entering the
Bell's inequality are in principle experimentally inaccessible, and hence
the author's claim is in principle not testable\textquotedblright .

There is a simple reply. To regard this as a criticism of the argument in
section 4 of \cite{1} is to misunderstand the structure of that argument.
That the probabilistic correlations predicted by quantum theory in the
scenario of the \textit{Gedankenexperiment} there described are not (in a
certain sense) testable is a feature of the argument, not a bug! That
argument proceeds by \textit{reductio ad absurdum}. It proves that a number
of plausible premises are in fact mutually inconsistent. If the proof is
valid then at least one of these premises must be false. The point of the
argument is to explore the consequences for objectivity of the assumption
that the probabilistic predictions of unitary quantum theory are correct,
including when the theory is applied to an entire isolated experimental
laboratory containing an experimenter performing a quantum measurement and
recording its outcome. Whether these predictions can be tested in the
scenario of the \textit{Gedankenexperiment} is simply irrelevant to the
validity of the argument.

Here is the sense in which the probabilistic correlation functions predicted
by quantum theory in the scenario of the \textit{Gedankenexperiment}
described in section 4 of \cite{1} are untestable. Frequency data of the
outcomes of every repetition of the sequence of single quantum measurements
by Alice, Bob, Carol and Dan could not all be tabulated together in a
single, localized space-time region within the world-tube of a single
spatially-localized experimenter. So no localized agent (not just Alice,
Bob, Carol or Dan) could compare the statistical correlation functions
computed from these data with the probabilistic correlation functions
predicted by quantum theory to see how well they match. I take it that this
is what is meant by the first sentence of the passage from \cite{2}
describing the \textquotedblleft first problem\textquotedblright\ with the
argument of \cite{1}:

\textquotedblleft The violation of the proposed Bell's inequality cannot,
not even in principle, be tested, because in no region of space-time are the
experimental data from which all the correlations functions can be extracted
available.\textquotedblright

Of course, the assumption (A) implies that all this experimental data is
present somewhere in the much larger space-time region in which are located
all the quantum measurements performed throughout the duration of the 
\textit{Gedankenexperiment}.

(A)\qquad Every quantum measurement has a definite (unique, objective)
physical outcome.

But, though (by assumption) present there, all this data is not
epistemically accessible to any individual experimenter. Any attempt by an
experimenter to access all this data would involve physical interactions
that would break the isolation of the experimenters' labs involved in the 
\textit{Gedankenexperiment} and disrupt the delicate sequence of operations
they perform in each repetition wiithin the \textit{Gedankenexperiment}.

That the violation of the proposed Bell's inequality cannot, not even in
principle, be tested raises no problem for the argument of \cite{1}.

The authors of \cite{2} attempt to raise a second problem that would be
faced by anyone wishing to remove the alleged first problem by
\textquotedblleft attributing an operational meaning to the computed
expressions for the [probabilistic] correlation functions\textquotedblright
. But this is no problem, since, having dismissed the first
\textquotedblleft problem\textquotedblright , anyone giving the argument of 
\cite{1} has no need to \textquotedblleft attribute an operational meaning
to the computed correlation functions\textquotedblright\ and should not wish
to do so.

As the authors of \cite{2} correctly note, if Alice and Bob were to try to
test their predictions and actually violate a Bell's inequality with data
they had collected, they would have to adapt the experimental protocol in
some way (which then may or may not in fact predict such violation). But the
validity of the argument of \cite{1} neither requires nor motivates any
change in the protocol there described. As it stands, that argument does not
require \textquotedblleft the standard assumptions for testing the violation
of Bell's inequalities namely `freedom of choice' and
`locality'\textquotedblright .

In a section entitled \textquotedblleft Analysis of the
protocol\textquotedblright\ the authors of \cite{2} discuss what they call
\textquotedblleft fundamental problems with the proposal in \cite{1}%
\textquotedblright . After repeating the (true, but irrelevant) claim of
untestability, they diagnose \textquotedblleft multiple
problems\textquotedblright\ stemming from the assumption that the
measurement of Carol and Dan are considered as unitary transformations. Now
it is a basic assumption of the entire \textit{reductio} argument that every
interaction may be viewed as proceeding in accordance with a unitary
transformation in the relevant total quantum state. The argument shows that
this assumption leads to conflict with assumption (A)--that was the point of
the argument, not a problem with it.

The first of the \textquotedblleft multiple problems\textquotedblright\
concerns an ambiguity that allegedly arises from the fact that certain
probabilistic correlation functions are computed from two different
reference frames. The authors of \cite{2} note that that there would be no
ambiguity if these correlation functions would refer to measurement data,
since these are reference frame independent. But the correlation functions 
\textit{do} refer to measurement data: in the correlation $E(a,d)$ they
refer to data collected in measurements by Alice and by Dan, while in the
correlation $E(b,c)$ they refer to data collected in measurements by Bob and
by Carol. In a single repetition within the \textit{Gedankenexperiment}
these data are collected in the regions of Figure 1 of \cite{1} there
labeled $U_{A}^{1},$\ $U_{D}^{2},$ $U_{B}^{2},$ $U_{C}^{1}$ respectively.
Assumption (A) guarantees the objectivity of this data, even while each of
their measurement interactions is assumed to correspond to the unitary
transformation used to label these regions of the figure.

The second of these \textquotedblleft problems\textquotedblright\ alleges
another kind of ambiguity, of a kind first alleged in \cite{3}. Alternative
ways of calculating $E(a,d)$ and $E(b,c)$ are proposed, in each case leading
to an alternative value for this quantity: specifically, evaluated in
Alice's reference frame $E(b,c)=0$ for all times rather than $-cos(b-c)$ as
claimed in \cite{1}. But this evaluation in Alice's frame is incorrect,
since it assumes that it is possible to assign a value to the outcome of a
measurement by Bob (Carol) even though Bob's (Carol's) register is in, or
returned to, its fixed pre-measurement state. The only correlations
considered in the argument of \cite{1} are correlations between the
measurement outcomes whose objective occurrence follows from assumption (A).
Each of these outcomes occurs in the localized space-time region where the
relevant measurement is performed. No outcome occurs in a region where a
register is in its pre-measurement state: and a measurement by Carol (Dan)
occurs when and where she (he) makes it, not in a space-time region where
Alice or Bob might have performed a measurement on Carol's (Dan's) lab to
try to inspect their outcomes. This second problem does not arise for the
reason the authors of \cite{2} themselves state:

\textquotedblleft If the four measurements would be identified with four
space-time points in which counts are registered, then the correlations
between these counts will be reference-frame independent and there would be
no problem.\textquotedblright\ The \textquotedblleft main
criticism\textquotedblright\ made in \cite{2} of the argument in section 4
of \cite{1} leaves that argument untouched, and, on careful examination, the
\textquotedblleft multiple problems\textquotedblright\ alleged in \cite{2}
turn out not to be problems at all.

\begin{acknowledgement}
This note was composed while I\ was a Fellow of the Stellenbosch Institute
for Advanced Studies: I am grateful to STIAS for their support.
\end{acknowledgement}

\end{document}